\documentclass[prl,twocolumn,superscriptaddress,showpacs]{revtex4}
\usepackage{amssymb}
\usepackage{epsfig}
\usepackage{amsmath}
\usepackage{amsfonts}

\begin{document}

\title{Triplet pairing due to spin-orbit-assisted electron-phonon coupling.}
\author{Vladimir I. Fal'ko}
\affiliation{Physics Department, Lancaster University, Lancaster, LA1 4YB, UK}
\author{Boris N. Narozhny}
\affiliation{The Abdus Salam ICTP, Strada Costiera 11, Trieste, I-34100, Italy}
\date{\today}

\begin{abstract}
We propose a microscopic mechanism for triplet pairing due to
spin-orbit-assisted electron interaction with optical phonons in a crystal
with a complex unit cell. Using two examples of electrons with symmetric
Fermi surfaces in crystals with either a cubic or a layered square lattice,
we show that spin-orbit-assisted electron-phonon coupling can, indeed,
generate triplet pairing and that, in each case, it predetermines the tensor
structure of a $p$-wave order parameter.
\end{abstract}

\maketitle

Spin-triplet superconductivity has been found in several materials including 
$Sr_{2}RuO_{4}$ \cite{ex1} and uranium-based compounds ($UPt_{3}$, $UBe_{13}$%
, $UGe_{2}$, $URuGe$) \cite{ex2,ex3,ex4,exURuGe}. While the phenomenology of
the triplet state is well known (it is similar to that of $p$%
-wave-superfluid $He^{3}$ \cite{leg,abm}), the microscopic origin of the
triplet pairing in each of these materials and its relation to the band
structure are the subject of intense investigation \cite%
{th1,Rice,Pickett,Kirkpatrick,ex1,ex2,ex3,ex4}, and the existing models \cite%
{th1,Rice,Pickett,Kirkpatrick} attribute the nature of the electron-electron
attraction in the triplet channel to the interaction via magnetic
fluctuations.

In this Letter, we show that a strong spin-orbit coupling in constituent
atoms of a metal with a complex unit cell can generate triplet pairing via
mechanism which involves optical phonons (instead of spin waves). The
attention \cite{th4} to spin-orbit coupling in the electron band structure
in the context of odd-parity pairing has been attracted by the observation
of superconductivity in the inversion-asymmetric $CePt_{3}Si$ \cite{ex5}. In
other known triplet superconductors, spin-orbit coupling in the band
structure can be ignored due to the inversion symmetry of their crystalline
lattice. However, in center-symmetric crystals with a multi-component unit
cell sublattice displacements may break inversion symmetry, so that atomic
spin-orbit coupling can give rise to an additional, spin-dependent
interaction between electrons and optical phonons. Such a
spin-orbit-assisted electron-phonon (e-ph) coupling arises from a
spin-dependent envelope between orbitals belonging to atoms from different
sublattices (mutially displaced by the vibration in the optical phonon
mode), and it may generate attraction in the spin-triplet Cooper channel via
the BCS-like mechanism.

Below, we demonstrate that the spin-orbit-assisted e-ph coupling does,
indeed, lead to spin-triplet $p$-wave pairing. The symmetry of the order
parameter of the resulting superconducting state is determined by the
particular crystalline lattice and the polarization properties of the
strongest-coupled optical phonon modes, which we show for two model
examples: of (i) a cubic lattice with two atoms in the unit cell and a
spherical Fermi surface and (ii) a layered structure with in-plane square
symmetry and a cylindrical Fermi surface.

For a cubic center-symmetric two-component lattice the electron spin
coupling to the inversion symmetry breaking sublattice displacement $\mathbf{%
w}$ can be described phenomenologically using the invariant \cite{footnoteSO}
\begin{equation}
\hat{\mathrm{V}}_{\mathrm{c}}=(\alpha /2)\left( \mathbf{w\cdot }\left[ \hat{%
\mathbf{p}}\times \hat{\mathbf{\sigma }}\right] +\left[ \hat{\mathbf{p}}%
\times \hat{\mathbf{\sigma }}\right] \mathbf{\cdot w}\right) .  \label{vc}
\end{equation}

\noindent\ In the second-quantized form, this can be represented as 
\begin{gather}
\hat{\mathcal{V}}=\sum\limits_{\mathbf{p},\mathbf{q},\zeta }V_{\mu \nu
}^{\zeta }(\mathbf{p,q})\hat{f}_{\mu \mathbf{p}+\frac{\mathbf{q}}{2}%
}^{\dagger }\hat{f}_{\nu \mathbf{p}-\frac{\mathbf{q}}{2}}\left( \hat{a}_{%
\mathbf{q}\zeta }+\hat{a}_{-\mathbf{q}\zeta }^{\dagger }\right) ,
\label{heph} \\
\hat{V}_{\mathrm{c}}^{\zeta }=\left( \alpha /\sqrt{2\omega _{\zeta }m_{\zeta
}}\right) \,\mathbf{l}_{\zeta }\mathbf{\cdot }\left[ \mathbf{p}\times 
\mathbf{\hat{\sigma}}\right] .  \notag
\end{gather}%
Here, $\hat{a}^{\dagger }(\hat{a})$ and $\hat{f}^{\dagger }(\hat{f})$ stand
for creation(annihilation) operators of phonons and electrons, $\mathbf{l}%
_{\zeta }$, $\omega _{\zeta }$ and $m_{\zeta }$ are the polarisation vector,
frequency and reduced mass of the optical phonon $\zeta $, indices $\mu $,$%
\nu $ identify the electron in the Kramers doublet for each momentum (spin
for a simple conduction band), and the parameter $\alpha $ is a material
constant. In a cubic crystal, we neglect a small difference between the LO
and TO phonon frequencies, thus $\omega _{\zeta }\approx \omega _{\mathrm{o}%
} $. In a less symmetric crystal, $\hat{V}^{\zeta }=\sum_{k=x,y,z}\hat{\sigma%
}^{k}A_{k}^{\zeta }(\mathbf{p},\mathbf{q})$, where $A_{k}^{\zeta }(\mathbf{p}%
,-\mathbf{q})=A_{k}^{\zeta }(\mathbf{p},\mathbf{q})$ and $A_{k}^{\zeta }(-%
\mathbf{p},\mathbf{q})=-A_{k}^{\zeta }(\mathbf{p},\mathbf{q})$, thus, $\hat{V%
}^{\zeta }(-\mathbf{p,q})=-\hat{V}^{\zeta }(\mathbf{p,q})$, $\hat{V}^{\zeta
}(\mathbf{p,-q})=\hat{V}^{\zeta }(\mathbf{p,q})$. For example, in a layered
crystal with a square lattice where we neglect the electron motion
perpendicular to the x-y planes, we take into account two invariants
analogous to Eq. (\ref{vc}), $\alpha _{\perp }w^{z}\left( p^{x}\hat{\sigma}%
^{y}-p^{y}\hat{\sigma}^{x}\right) $ and $\alpha _{\parallel }\hat{\sigma}%
^{z}\left( w^{x}p^{y}-w^{y}p^{x}\right) $, to distinguish between the
electron coupling to optical phonons polarized perpendicular ($\zeta
=z;\omega _{z}=\omega _{\perp }$) and in the plane of the layers ($\zeta
=x,y;\omega _{x,y}=\omega _{\parallel }$): 
\begin{equation}
\hat{V}_{\mathrm{sq}}^{z}=\frac{\alpha _{\perp }\left( p^{x}\hat{\sigma}%
^{y}-p^{y}\hat{\sigma}^{x}\right) }{\sqrt{2\omega _{\perp }m_{\perp }}},\;%
\hat{V}_{\mathrm{sq}}^{x(y)}=\frac{\pm \alpha _{\parallel }p^{y(x)}\hat{%
\sigma}^{z}}{\sqrt{2\omega _{\parallel }m_{\parallel }}}.  \label{layered}
\end{equation}

A spin-dependent e-ph coupling, Eq.~(\ref{heph}) and a virtual polarisation
of the crystalline unit cell \cite{txb} result in the effective
spin-dependent interaction between electrons on the opposite sides of the
Fermi surface,%
\begin{eqnarray}
\mathcal{U} &=&-\sum_{\mathbf{p}_{1},\mathbf{p}_{2}}M_{\gamma \delta
}^{\alpha \beta }\hat{f}_{\alpha \mathbf{p}_{1}}^{\dagger }\hat{f}_{\gamma -%
\mathbf{p}_{1}}^{\dagger }\hat{f}_{\delta -\mathbf{p}_{2}}\hat{f}_{\beta 
\mathbf{p}_{2}},  \label{h} \\
M_{\gamma \delta }^{\alpha \beta } &=&\sum_{\zeta }\frac{\omega _{\zeta
}V_{\alpha \beta }^{\zeta }(\mathbf{p}_{1}+\mathbf{p}_{2})V_{\gamma \delta
}^{\zeta }(-\mathbf{p}_{2}-\mathbf{p}_{1})}{\omega _{\zeta }^{2}-\left(
\varepsilon _{\mathbf{p}_{2}}-\varepsilon _{\mathbf{p}_{1}}\right) ^{2}}.
\label{m}
\end{eqnarray}%
In Eq. (\ref{m}), we neglect the dispersion of optical phonon modes, $\omega
_{\zeta }(q)\approx \omega _{\zeta }$ and take into account the
quasiparticle states close to the Fermi surface ($p=|\mathbf{p}|\approx
p_{F} $), with $\left\vert \varepsilon _{\mathbf{p}_{i}}\right\vert <\omega
_{\zeta }$, where $\varepsilon _{\mathbf{p}}$ is the quasiparticle energy
calculated with respect to the Fermi level. The effective electron-electron
(e-e) interaction, $\hat{\mathcal{U}}$ is spin- and momentum-dependent,
therefore, it may allow for attraction in the spin-triplet channel and odd
pairing.

To determine what triplet phase can be formed in a metal with the Cooper
interaction in Eqs.(\ref{h},\ref{m}), one needs to identify
particle-particle channels in which the electron-electron interaction is
attractive. The classification of pairing channels and the resulting triplet
order parameter is usually based upon the symmetry of the electronic
Hamiltonian \cite{leg}. Following a procedure developed for $p$-wave
superfluidity in $He^{3}$ \cite{leg,abm}, we determine operators of singlet (%
$\hat{b}$) and triplet ($\hat{t}_{s}$) electron pairs, 
\begin{equation}
\hat{b}(\mathbf{p})=i\sigma _{\mu \nu }^{y}\hat{f}_{\mu -\mathbf{p}}\hat{f}%
_{\nu \mathbf{p}};\,\hat{t}_{s}(\mathbf{p})=i(\sigma ^{y}\sigma ^{s})_{\mu
\nu }\hat{f}_{\mu -\mathbf{p}}\hat{f}_{\nu \mathbf{p}}.  \label{bop}
\end{equation}%
In terms of these operators \cite{footnote2}, the Hamiltonian $\mathcal{U}$
in Eq. (\ref{h}) can be written down in a separable form, 
\begin{equation}
\mathcal{U}=-\sum_{\mathbf{p}_{1}\mathbf{p}_{2}}\left[
T^{s_{1}s_{2}}t_{s_{1}}^{\dagger }(\mathbf{p}_{1})t_{s_{2}}(\mathbf{p}_{2})+C%
\hat{b}^{\dagger }(\mathbf{p}_{1})\hat{b}(\mathbf{p}_{2})\right] ,
\label{mt}
\end{equation}%
which includes the interaction both in the singlet channel \cite{footnote3},
with the constant $C=M_{\gamma \delta }^{\alpha \beta }\sigma _{\gamma
\alpha }^{y}\sigma _{\beta \delta }^{y}$, and in the triplet channel, with%
\begin{equation}
T^{s_{1}s_{2}}=M_{\gamma \delta }^{\alpha \beta }\left( \hat{\sigma}^{y}\hat{%
\sigma}^{s_{1}}\right) _{\gamma \alpha }\left( \hat{\sigma}^{s_{2}}\sigma
^{y}\right) _{\beta \delta }.  \label{Tsl}
\end{equation}

Below, we limit the analysis to the case of $p$-wave pairing (excluding
higher-order angular harmonics) described by the tensor order parameter,

\begin{equation}
B_{ms}=\sum_{\mathbf{p}}n_{m}\langle \hat{t}_{s}(\mathbf{p})\rangle ;\quad 
\mathbf{n}=\frac{\mathbf{p}}{p},  \label{B}
\end{equation}%
and focus it on two model examples: (i) a 3D metal with a cubic lattice and
a spherical Fermi surface and (ii) a layered material with a square in-plane
lattice and a cylindrical Fermi surface. In such symmetric systems, triplet
pairs can be characterised by well-defined quantum number related to
irreducible tensor representations of the corresponding symmetry group.

(i). In the case of a model\textit{\ metal with a cubic lattice and a
spherical Fermi surface} the symmetry is the full rotational group SO$_{3}$.
The involment of the spin-orbit coupling in the formation of the pair
interaction assumes that the spin and orbital degrees of freedom of the
Cooper pair are not independent, so that its total spin $S=1$ and orbital
angular momentum, $L=1$ sum into the total angular momentum, $J=0,1,2$. The
tensor structure of the order parameter for these values of $J$ is the
following: $B_{ms}^{J=0}=\delta ^{ms}De^{i\phi }$, where $B$ is a scalar; $%
B_{ms}^{J=1}=\epsilon ^{msl}D_{l}$, where $\mathbf{D}$ is a complex vector;
and $B_{ms}^{J=2}$ is a traceless symmetric tensor.

Thus, decompose e-e interaciton in Eq. (\ref{mt}) into three channels
corresponding to different angular momenta of the pair. Using the relation
of the matrix $M_{\gamma \delta }^{\alpha \beta }$ with the e-ph coupling $%
\hat{V}_{\mathrm{c}}^{\zeta }$ in Eq. (\ref{heph}), we find that 
\begin{eqnarray*}
&&T^{s_{1}s_{2}}=g_{\mathrm{so}}\sum%
\limits_{m_{1},m_{2}}n_{1}^{m_{1}}n_{2}^{m_{2}}G_{m_{1}m_{2}}^{s_{1}s_{2}};
\\
&&G_{m_{1}m_{2}}^{s_{1}s_{2}}=\varkappa _{0}\delta ^{m_{1}s_{1}}\delta
^{m_{2}s_{2}}+\varkappa _{1}\epsilon ^{m_{1}s_{1}l}\epsilon ^{m_{2}s_{2}l}+
\\
&&\quad +\varkappa _{2}\left[ \delta ^{m_{1}m_{2}}\delta
^{s_{1}s_{2}}+\delta ^{m_{1}s_{2}}\delta ^{m_{2}s_{1}}-\frac{2}{3}\delta
^{m_{1}s_{1}}\delta ^{m_{2}s_{2}}\right] .
\end{eqnarray*}%
\noindent The first term in the matrix $G_{m_{1}m_{2}}^{s_{1}s_{2}}$ is a
product of projectors onto the $J=0$ paired state; the second projects onto
the states with $J=1$, whereas the last term selects only the states with $%
J=2$. Interaction in each channel is described by the corresponding
constant, 
\begin{equation}
g_{J}=\varkappa _{J}g_{\mathrm{so}},\;\,g_{\mathrm{so}}\sim \left( \alpha
/\omega _{\mathrm{o}}\right) ^{2}p_{F}^{2}/2m_{\mathrm{o}},  \label{gJ}
\end{equation}%
where the scale of the interaction depends on the strength of the original
electron-phonon interaction for electrons near the Fermi surface, frequency $%
\omega _{\mathrm{o}}$ and reduced mass $m_{\mathrm{o}}$ of the optical
phonon (here we disregard a small difference between longitudinal and
transverse optical phonon modes in a cubic crystal). \ Interaction in
channels with different $J$'s\ differs by the numerical factors, 
\begin{equation}
\varkappa _{0}=-\frac{4}{3};\varkappa _{1}=\frac{1}{2};\varkappa _{2}=-\frac{%
1}{2}.  \label{kappas}
\end{equation}

According to Eq. (\ref{mt}), the positive sign of $g_{J}$ corresponds to
attraction. Since only one of the effective coupling constants $g_{J}=g_{%
\mathrm{so}}\varkappa _{J}$ is positive, we conclude that in the cubic
system Cooper interaction due to the spin-orbit-assisted e-ph coupling leads
to the condensate of pairs with the total angular momentum $J=1$.

By selecting a pairing channel with $J=1$ we do not fully determine the
order parameter $B_{ms}$ in the lowest energy triplet state, yet. In $%
B_{ms}^{J=1}=\epsilon ^{msl}D_{l}$, $\mathbf{D=}De^{i\phi }\left( \mathbf{l}%
+i\mathbf{l}^{\prime }\right) /\sqrt{\mathbf{l}^{2}+\left( \mathbf{l}%
^{\prime }\right) ^{2}}$ is a complex vector \cite{leg,fsh} parametrized
using two perpendicular vectors, $\mathbf{l}$ and $\mathbf{l}^{\prime }$, $%
\mathbf{l}^{2}\geq \left( \mathbf{l}^{\prime }\right) ^{2}$. To find a
favourable state, we minimize the condensate energy, 
\begin{equation}
\mathcal{E}=\sum\limits_{\mathbf{p}\mu }\left( |\varepsilon _{\mathbf{p}%
}|-E_{\mathbf{p}\mu }\right) +\sum\limits_{ms}g\left\vert B_{ms}\right\vert
^{2},  \label{gse}
\end{equation}%
as a function of $\mathbf{D}$, which includes finding the relative size of
two vectors $\mathbf{l}$ and $\mathbf{l}^{\prime }$ and the absolute value
of $D$. The expression for $\mathcal{E}$ in Eq. (\ref{gse}) can be formally
obtained using the Hubbard-Stratonovich transformation which splits the
electron-electron interaction in Eq. \ref{mt} using the field $B_{ms}$ and,
then, by integrating out fermionic degrees of freedom. \ Equation (\ref{B})
is the self-consistency equation for the minimum of $\mathcal{E}$, and 
\begin{equation}
E_{\mathbf{p}\mu }=\sqrt{\varepsilon _{\mathbf{p}}^{2}+|\mathbf{d}_{\mathbf{p%
}}|^{2}-\mu \;\left\vert \left[ \mathbf{d}_{\mathbf{p}}\times \mathbf{d}_{%
\mathbf{p}}^{\ast }\right] \right\vert }  \label{es}
\end{equation}%
describes the spectrum of excitations \cite{leg,fsh} in the phase with order
parameter $B_{ms}$, where three components of the complex vector $\mathbf{d}%
_{\mathbf{p}}$ are defined as 
\begin{equation}
d_{\mathbf{p}}^{s}=g\sum\limits_{m}n_{m}B_{ms}.  \label{ds}
\end{equation}

\noindent The vector product $i\mathbf{d}_{\mathbf{p}}\times \mathbf{d}_{%
\mathbf{p}}^{\ast }$ determines the spin quantization axis and spin
splitting for Bogolyubov quasiparticles for each direction $\mathbf{n}=%
\mathbf{p}/p$ on the Fermi sphere and $\mu =\pm 1$ denotes the polarisation
of the quasiparticle spin with respect to such an axis. In Eqs. (\ref{gse},%
\ref{es}), $\varepsilon _{\mathbf{p}}$ is the quasiparticle energy in the
normal state calculated with respect to the Fermi level, and $g$ is the
interaction constant in the attraction channel (here, $g=g_{1}\equiv
\varkappa _{1}g_{\mathrm{so}}$).

In the context of pairing with $J=1$ and $B_{ms}^{J=1}=\epsilon ^{msl}D_{l}$%
, two possible extrema of the function $\mathcal{E}$ can be found using the
analysis of the degeneracy space of the order parameter \cite{fsh}. For
fixed $|\mathbf{D}|=D$, $\mathcal{E}(\mathbf{D})$ takes extremal values on
vectors $\mathbf{D}$ that cover minimal surfaces under the application of SO$%
_{3}$ rotations. Two such possibilities are $\mathbf{D}_{0}\mathbf{=l}%
e^{i\phi }D$ where $\mathbf{l}$ is a real unit vector (that's, $\mathbf{l}%
^{\prime }=0$), and $\mathbf{D}_{1}\mathbf{=}\frac{\mathbf{l}+i\mathbf{l}%
^{\prime }}{\sqrt{2}}e^{i\phi }D$, where $\mathbf{l}$ and $\mathbf{l}%
^{\prime }$ are two perpendicular unit vectors. For each of these two cases
we find the value of $D$ by minimizing the ground state energy $\mathcal{E}$
in Eq. (\ref{gse}). The results \cite{DOS} are shown in Table~\ref{ecu}
indicating that the phase with $\mathbf{D}_{0}\mathbf{=l}e^{i\phi }D$ and an
anisotropic gap vanishing along the direction $\mathbf{l}$ (since $\mathbf{d}%
=gD[\mathbf{n}\times \mathbf{l}]$) is energetically favorable.

\begin{table}[tbp]
\caption{$J=1$ phases due to spin-orbit-assisted electron-phonon coupling in
a metal with spherical Fermi surface and cubic lattice.}
\label{ecu}
\begin{ruledtabular}
\begin{tabular}{ccc}
Vector  ${\bf D}$ & mean-field $D$ & Energy, $\mathcal{E}$\\
${\bf l}e^{i\phi}D$ & $D=(2\omega_{{\rm o}}/g)e^{-3/g\nu}$ & 
  $\mathcal{E}=-\nu g^2 D^2/3$ \\
$\frac{{\bf l}+{\bf l'}}{\sqrt{2}}e^{i\phi}D$ & $D=(\sqrt{2}\omega_{{\rm o}}/g)e^{-6/g\nu}$ & 
     $\mathcal{E}=-\nu g^2 D^2/6$ \\
\end{tabular}
\end{ruledtabular}
\end{table}

Although the example of a cubic crystal with highly symmetric Fermi surface
may be too abstract, nevertheless, it illustrates that the
spin-orbit-assisted e-ph coupling can generate triplet pairing between
electrons and that the suggested pairing mechanism is highly prescriptive
for the structure of the resulting order parameter.

(ii). In a \textit{layered material with a square-symmetric lattice}, the
spin-orbit-assisted e-ph coupling is also able to generate triplet pairing
with definite phases. For simplicity (and with the reference to an almost
cylindrical $\gamma $-sheet in the Fermi surface of $Sr_{2}RuO_{4}$ \cite%
{ex1}) we consider below a metal with a cylindrical Fermi surface. Similarly
to the BCS theory, the effective attraction $\mathcal{U}$ in Eqs. (\ref{h},%
\ref{mt}) is point-like, so that only electrons from the same layer can form
a pair. Thus, we shall consider only two-dimensional order parameters (the
interlayer tunneling can be treated using the Lawrence-Doniach approach \cite%
{txb}). In this case, triplet \textit{p}-wave pairs can be characterized by
the component $J_{z}$ of the total angular momentum in the direction
perpendicular to the layers, which is the sum of the projection $L_{z}=\pm 1$
of their orbital angular momentum and $S_{z}=0,\pm 1$ of the pair spin. For $%
J_{z}=\pm 1$ and $J_{z}=\pm 2$, the quantum number $J_{z}$ fully determines
the order parameter, whereas for $J_{z}=0$, pairs should be additionally
classified according to whether they are symmetric ($s$) or antisymmetric ($%
a $) with respect to in-plane $x\rightleftarrows y$ reflection, for which we
reserve notations $J_{z}=0^{s}$ and $J_{z}=0^{a}$, respectively. The second
line in Table \ref{ps} summarizes tensor structure, $%
B_{ms}=P_{ms}^{J_{z}}e^{i\phi }D$ of the order parameter for phases with $%
J_{z}=0^{s}$,$0^{a}$,$\pm 1$,$\pm 2$, where matrices $P_{ms}^{J_{z}}$ form
the orthogonal basis which realise irreducible representations of $J_{z}$,
and where $D^{2}=\sum_{ms}|B_{ms}|^{2}$.

Using matrices $P^{J_{z}}$ the effective interaction $\hat{\mathcal{U}}$ can
be written down in the separable form, Eq. (\ref{mt}) with

\begin{gather*}
T^{s_{1}s_{2}}=\sum\limits_{m_{1}m_{2}J_{z}}(\varkappa _{J_{z}}g_{\mathrm{so}%
}^{\perp }+\tilde{\varkappa}_{J_{z}}g_{\mathrm{so}}^{\Vert
})n_{1}^{m_{1}}n_{2}^{m_{2}}P_{s_{1}m_{1}}^{J_{z}}P_{s_{2}m_{2}}^{J_{z}\;%
\ast }, \\
g_{\mathrm{so}}^{\zeta }\sim \left( \alpha _{\zeta }/\omega _{\zeta }\right)
^{2}p_{F}^{2}/2m_{\zeta },\;\;\zeta =\perp ,\Vert ,
\end{gather*}%
which is determined by couplings (\ref{layered}) to optical phonons
polarized across ($\zeta =\perp $) and parallel ($\zeta =\Vert $) to the
layer. The values of coefficients $\varkappa _{J_{z}}$ and $\tilde{\varkappa}%
_{J_{z}}$ are listed in Table \ref{ps}, which shows that positive coupling
constants (determining the attraction in the corresponding Cooper channel)
may appear only for electron pairs in the following channels: (a) $%
J_{z}=0^{a}$ where $g_{0^{a}}=\varkappa _{0^{a}}g_{\mathrm{so}}^{\perp }+%
\tilde{\varkappa}_{0^{a}}g_{\mathrm{so}}^{\Vert }=2g_{\mathrm{so}}^{\perp
}-g_{\mathrm{so}}^{\Vert }$, and (b) $J_{z}=\pm 1$ where $g_{\pm
1}=\varkappa _{\pm 1}g_{\mathrm{so}}^{\perp }+\tilde{\varkappa}_{\pm 1}g_{%
\mathrm{so}}^{\Vert }=g_{\mathrm{so}}^{\Vert }-g_{\mathrm{so}}^{\perp }$.
This indicates that, within the proposed mechanism, only two triplet phases
are possible.

To determine which phase is realized we need to compare the relative
strength of the electron coupling to the in- and out-of-plane sublattice
vibrations. If $g_{\mathrm{so}}^{\Vert }<\frac{3}{2}g_{\mathrm{so}}^{\perp }$%
, so\ that $g_{0^{a}}>g_{\pm 1}$, the ground state with $J_{z}=0^{a}$ is
realized. This phase is characterized by the vector order parameter $\mathbf{%
d}$ and an isotropic gap, 
\begin{equation}
\mathbf{d}_{0^{a}}=\frac{gD}{\sqrt{2}}\left( 
\begin{array}{c}
n^{y} \\ 
-n^{x} \\ 
0%
\end{array}%
\right) \;;\;\;\;%
\begin{array}{c}
E_{\mathbf{p}}=\sqrt{\varepsilon _{\mathbf{p}}^{2}+\Delta ^{2}}, \\ 
\Delta =2\omega _{\bot }e^{-2/\nu g}, \\ 
g=2g_{\mathrm{so}}^{\perp }-g_{\mathrm{so}}^{\Vert }.%
\end{array}
\label{phase1}
\end{equation}%
Here \cite{DOS}, $D=2^{3/2}(\omega _{\bot }/g)e^{-2/\nu g}$ \ was found from
minimizing $\mathcal{E}$ in Eq. (\ref{gse}) with respect to $D$.

\begin{widetext}

\begin{table}
\caption{\label{ps} Triplet pairing order parameter for electrons with cylindrical Fermi surface in a layered material with square lattice and the interaciton constants in the corresponding $J_z$ channels due to the spin-orbit-assisted e-ph coupling.}
\begin{ruledtabular}
\begin{tabular}{cccc}
$J_z=\pm 2$ & $J_z=\pm 1$ & $J_z=0^s$ & $J_z=0^a$ \\
& (ABM) & & \\
$P_{sm}^{\pm 2}=
 \frac{1}{2}\left( \delta^{mx}\pm i\delta^{my} \right)
 \left( \delta^{sx}\pm i\delta^{sy} \right)$ & 
$P_{sm}^{\pm 1}=
 \frac{1}{\sqrt{2}}
 \left( \delta^{mx}\pm i\delta^{my} \right) \delta^{sz}$ & 
$P_{sm}^{0^s}=
 \frac{1}{\sqrt{2}} \left( \delta^{mx}\delta^{sx} +
 \delta^{my} \delta^{sy} \right)$ &
$P_{sm}^{0^a}=
 \frac{1}{\sqrt{2}} \left( \delta^{my}\delta^{sx} -
 \delta^{mx} \delta^{sy} \right)$ \\
$\varkappa_{\pm 2} = 0 $ & $\varkappa_{\pm 1} =-1$  & 
     $\varkappa_{0^s} = -2$ & $\varkappa_{0^a} = 2$ \\
$\tilde\varkappa_{\pm 2} = -1$ & $\tilde\varkappa_{\pm 1} = 1$  & 
     $\tilde\varkappa_{0^s} = -1$ & $\tilde\varkappa_{0^a} = -1$ \\
\end{tabular}
\end{ruledtabular}
\end{table}
\end{widetext}

If $g_{\mathrm{so}}^{\Vert }>\frac{3}{2}g_{\mathrm{so}}^{\perp }$, so that $%
g_{\pm 1}>g_{0^{a}}$ (in terms of microscopic parameters, for a stronger
coupling\ to the in-plane sublattice displacement, $\alpha _{\Vert }/(\sqrt{%
m_{\Vert }}\omega _{\Vert })>\sqrt{\frac{3}{2}}\alpha _{\perp }/(\sqrt{%
m_{\perp }}\omega _{\perp })$), then the ground state belongs to the phase
with $J_{z}=\pm 1$ which is similar to the Anderson-Brinkmann-Morel (ABM)
phase discussed in the theory \cite{abm,txb} of superfluid $He^{3}$. \ This
phase also has a circularly isotropic gap: 
\begin{equation}
\mathbf{d}_{\pm 1}=\frac{gD}{\sqrt{2}}\left( 
\begin{array}{c}
0 \\ 
0 \\ 
n^{x}\pm in^{y}%
\end{array}%
\right) \;;\;\;\;%
\begin{array}{c}
E_{\mathbf{p}}=\sqrt{\varepsilon _{\mathbf{p}}^{2}+\Delta ^{2}}, \\ 
\Delta =2\omega _{\Vert }e^{-2/\nu g}, \\ 
g=g_{\mathrm{so}}^{\Vert }-g_{\mathrm{so}}^{\perp }.%
\end{array}
\label{phase2}
\end{equation}

In conclusion, we have shown that the spin-orbit-assisted e-ph coupling can
generate triplet pairing with a definite tensor structure of the $p$-wave
order parameter in each type of a crystal. For example, in a layered metal
where conducting planes have square lattice structure, the proposed
mechanism of triplet pairing leads to two possible phases in Eqs. (\ref%
{phase1}) and (\ref{phase2}), one of which (ABM) is actually considered \cite%
{ex1} as the most plausible candidate for the triplet order parameter in
superconditing $Sr_{2}RuO_{4}$ that has the layered perovskite structure
with body-centred tetragonal group symmetry. On the one hand, the latter
coincidence may be purely accidental, so that a detailed analysis of the
phonon spectrum and the spin-orbit-assisted e-ph coupling strength in each
particular material based upon first-principle calculations is needed. On
the other hand, the proposed pairing mechanism suggests a n isotope effect
of the usual sign in the triplet superconductivity materials (for example,
using oxigen isotopes in $Sr_{2}RuO_{4}$, as reported in Ref. \cite%
{isotope,isotopeRem}), which would manifest the involvement of e-ph coupling
in the formation of triplet pairing.

Authors thank A.Mackenzie and E.Tosatti for discussions. This work was
supported by EPSRC and EC NMP2-CT2003-505587 'SFINX'.


\begin{thebibliography}{99}
\bibitem{footnoteSO} We symmetrize $\hat{\mathrm{V}}_{\mathrm{c}}$ to take
into account non-commutation of the momentum $\hat{\mathbf{p}}$ and the
coordinate-dependent $\mathbf{w}(\mathbf{x})$.
\bibitem{footnote2} The singlet operator $\hat{b}$ is an even function of $%
\mathbf{p}$, while the triplet $\hat{t}_{s}$ are odd. Thus, the e-ph
coupling generates an effective electron-electron interaction in the triplet
channel, due to the assymmetry $\hat{V}^{\zeta }(-\mathbf{p,q})=-\hat{V}%
^{\zeta }(\mathbf{p,q})$ and $\hat{V}^{\zeta }(\mathbf{p,-q})=\hat{V}^{\zeta
}(\mathbf{p,q})$ of its spin-orbit-assisted part. The latter property is
specific to optical phonons. The acoustic e-ph coupling has a different
symmetry, $\hat{V}^{\zeta }(\mathbf{p,-q})=-\hat{V}^{\zeta }(\mathbf{p,q})$
and cannot generate the effective interaction in the $p$-wave channel.
\bibitem{footnote3} $S$-wave interaction in Eq. (\ref{mt}) slightly changes
the usual BCS coupling. This may be ignored if the latter is less than the
Coulomb repulsion. In the $p$-wave channel Coulomb interaction is
suppressed, due to screening.
\bibitem{DOS} $\nu $ is the Fermi density of states in one spin component.
\bibitem{isotopeRem} The sign of the isotope effect observed in Ref. \cite%
{isotope} depended on the quality of the structure and it is not clear
whether it was related to the change of the frequency of the optical phonon
mode upon the isotopic substitution. The inverse sign of the isotope effect
in the purest samples in Ref. \cite{isotope} may have been the result of
pair breaking by the annealing-induced disorder and differently diffusing
light and heavy isotopes.
\end{thebibliography}
\end{document}